\documentclass[12pt]{iopart}
\usepackage{graphicx}
\begin{document}

\title[Photoemission studies of Fe(Se,Te)]{Photoemission studies of the near 
E$_F$ spectral weight shifts in
FeSe$_{1-x}$Te$_{x}$ superconductor\\}

\author{P. Mishra$^{1}$, H. Lohani$^{1}$, R.A. Zargar$^{2}$, V.P.S. Awana$^{3}$, B.R. Sekhar$^{1}$}

\address{$^{1}$ Institute of Physics, Sachivalaya Marg, Bhubaneswar 751005, India.}
\address{$^{2}$ Department of Physics, Jamia Millia Islamia, New Delhi-110025, India}
\address{$^{3}$ CSIR-National Physical Laboratory, Dr. K. S. Krishnan Marg, New Delhi-110012, India}
\ead{sekhar@iopb.res.in}
\vspace{10pt}
\begin{indented}
\item[]July 2014
\end{indented}

\begin{abstract}
Our valence band photoelectron spectroscopic studies show a temperature 
dependent spectral weight transfer near the Fermi level in the Fe-based 
superconductor FeSe$_{1-x}$Te$_{x}$. Using theoretical band structure 
calculations we have shown that the weight transfer is due to the temperature 
induced changes in the Fe(Se,Te)$_{4}$ tetrahedra. These structural changes 
lead to shifts in the electron occupancy from the xz/yz and x$^{2}$-y$^{2}$ 
orbitals to the 3z$^{2}$-r$^{2}$ orbitals indicating a temperature induced 
crossover from a metallic state to an Orbital Selective Mott (OSM) Phase. Our 
study presents the observation of a temperature induced crossover to a 
low temperature OSM phase in the family of Fe chalcogenides.
\end{abstract}

%
\vspace{2pc}
\noindent{\it Keywords}:  UV photoemission spectroscopy, superconductor, electronic structure, pseudogap regime 
%
%
\maketitle
%
%

\section{Introduction}
The unconventional superconductivity\cite{hanaguri} in iron based 
compounds\cite{kamihara,yeh} has attracted much attention due to its 
importance in elucidating a consolidated understanding of superconductivity 
in general. Unlike the cuprate High T$_c$ materials, superconductivity in 
these compounds involves some exotic interplay of structural and magnetic 
degrees of freedom. Although among them, Fe(Se,Te) is rather simple from a 
structural point of view, it shows a strong bearing of the spin fluctuations 
and structural and magnetic disorder on its superconducting properties. 
Despite, a number of reports addressing many of these issues, the roles of 
electron-phonon coupling, spin density wave states, quasiparticles etc. are 
still under intense debate.

The structure of FeSe$_{1-x}$Te$_{x}$ compounds consists of a stacking of 
edge sharing Fe(Se,Te)$_{4}$ tetrahedra without any spacer layer in 
between\cite{yeh}. The parent compound FeTe is a non superconductor 
and exhibits a monoclinic structure at low temperature. It also shows an 
antiferromagnetic spin density wave (SDW) transition at 70 K\cite{li}. The 
other parent, FeSe is a superconductor with a T$_{c}$ of 8 K\cite{hsu}. 
Substitution of Se at the Te site enhances the T$_c$ to a maximum of 15 K for 
x = 0.5\cite{yeh}. This enhancement is reported to be linked to the local 
structural symmetry breaking\cite{joseph} and the degree of disorder caused 
by the smaller ionic radius of Se\cite{yeh}, which is counter intuitive as 
disorder is expected to reduce the superconducting transition 
temperature\cite{chand}. Furthermore, the electronic properties of these 
superconductors with their moderate electron correlations, are controlled 
mostly by the competing inter- and intra- orbital interactions near the E$_F$ 
region. It has been pointed out\cite{medici} that these interactions generate 
an Orbital Selective Mott Phase (OSMP) in which electrons in some orbitals 
are Mott localized while others remain itinerant. Earlier, such an OSMP was 
identified in Ca$_{2-x}$Sr$_{x}$RuO$_{4}$ \cite{neupane, anisimov}. Recently, 
a doping dependent OSMP was proposed to be present in Fe(Se,Te) by {\it Craco 
et al.} and {\it Aichhorn et al.}\cite{craco,aichhorn} from a theoretical 
point of view.

Electron spectroscopic studies, particularly using ultra-violet photoelectron 
spectroscopy have shown that the near E$_{F}$ electronic states in the 
Fe(Se,Te) compositions are dominated by the Fe 3d and chalcogen p 
states\cite{yoshida,yokoya}. Changes in the Fe 3d - Se 4p/Te 5p correlation strength 
with doping or temperature lead to noticeable spectral weight shifts in the 
near E$_F$ states which could be intimately related to the superconducting 
properties of these materials. Although some of the earlier 
studies\cite{zhang,lin,okazaki} have reported such spectral weight shifts and 
identified the formation of a pseudogap thereby, their origins and nature  
are still not clear. Further, the correlation between the formation of the 
pseudogap and disorder also has not been addressed. This study shows that the 
temperature dependent normal state pseudogap is intimately related to the 
insulating behavior originating from the multi-orbital correlation and the 
Hund's coupling. Based on our theoretical calculations we have shown that 
the formation of such a pseudogap could be a signature of the temperature 
induced Orbital Selective Mott Transition (OSMT).

\section{EXPERIMENTAL}

Poly crystalline samples of FeSe$_{1-x}$Te$_{x}$ (x = 1, 0.5, 0) were 
synthesized via solid state reaction route described elsewhere\cite{awana}. 
The stoichiometric compositions of the samples studied using 
X-ray diffraction and resistivity measurements have been published 
earlier\cite{zargar}. It should be noted that the samples contain no 
excess Fe and are of single phase nature. Angle integrated ultraviolet 
photoemission measurements were 
performed by using an ultra high vacuum system equipped with a high intensity 
vacuum-ultraviolet source and a hemispherical electron 
energy analyzer (SCIENTA R3000). At the He $I$ ($h$ $\nu$ = $21.2$ eV) line, 
the photon flux was of the order of $10^{16}$ photons/sec/steradian with a 
beam spot of $2$ mm diameter. Fermi energies for all measurements were 
calibrated by using a freshly evaporated Ag film on a sample holder. The 
total energy resolution, estimated from the width of the Fermi edge, was 
about $27$ meV for He $I$ excitation. All the photoemission measurements were 
performed inside the analysis chamber under a base vacuum of $\sim$ $5.0$ 
$\times$ $10^{-11}$ mbar. The polycrystalline samples were repeatedly scraped 
using a diamond file inside the preparation chamber with a base vacuum of 
$\sim$ $5.0$ $\times$ $10^{-11}$ mbar and the spectra were taken within $1$ 
hour, so as to avoid any surface degradation. All measurements were repeated 
many times to ensure the reproducibility of the spectra. For the temperature 
dependent measurements, the samples were cooled by pumping liquid nitrogen 
through the sample manipulator fitted with a cryostat. Sample temperatures 
were measured using a silicon diode sensor touching the bottom of the 
stainless steel sample plate. The low temperature photoemission measurements 
at $77$ K were performed immediately after cleaning the sample surfaces 
followed by the room temperature measurements. 

In order to understand the observed changes in the near E$_F$ spectral features 
we used TBLMTO-ASA\cite{andersen} calculations employing scalar relativistic 
corrections including combined correction term and Langreth-Meh-Hu gradient 
corrected von Barth Hedin parametrized energy and potential. The experimental 
lattice parameters at 300K were used in the calculations\cite{zargar}. The 
correlation effects of the Fe-d orbitals were taken into account by using LDA+U 
formulism with J as 0.9 eV\cite{aichhorn}, and U as 3.5 eV for FeTe and 4.0 eV 
for FeSe\cite{miyake}. For the FeSe$_{0.5}$Te$_{0.5}$ U was taken as 3.8 eV, 
a value intermediate between those of FeTe and FeSe.

\section{RESULTS AND DISCUSSION}

Figure 1(b) shows the valence band spectra of FeSe$_{1-x}$Te$_{x}$ (x = 1, 
0.5, 0) samples taken at He I photon energy. The spectral features marked A 
and B, positioned at 0.5 eV and 2 eV respectively, originate from the Fe 3d 
states. Peak C at 4 eV is due to the hybridized Fe 3d - Se 4p/Te 5p states 
while D at 6 eV corresponds to Se 4p/Te 5p states. Our calculations based on 
the TBLMTO-ASA (fig 2(a)) and also other's calculations\cite{miyake,subedi} 
conform to these assignments. The sharp feature A seen in the case of FeSe 
transforms into a broadened one for FeTe. Feature C gets broadened and shifts 
to lower binding energy with increasing x. Further, feature D also shifts to 
lower binding energy with doping.

In order to see the finer changes in the near E$_F$ electronic structure we 
have taken a set of high resolution spectra of this region. Figure 1(c) 
depicts the near E$_{F}$ valence band spectra of FeSe$_{1-x}$Te$_{x}$ at two 
different temperatures. Black and red spectra correspond to 300 K and 77 K 
respectively. The feature A, originating from Fe 3d states shows an 
enhancement in its intensity as the temperature is lowered. This enhancement 
is prominent in case of FeSe$_{0.5}$Te$_{0.5}$. We will discuss this point 
in the following paragraph. {\it Yokoya et al.}\cite{yokoya} have earlier shown that the 
feature A consists of two features, A$^{\prime}$ and A$^{\prime\prime}$ 
(see Fig. 1(c)). As we go from FeTe to FeSe, the energy separation between 
these two features keep decreasing and intensity of A$^{\prime\prime}$ increases.
Thus, the doublet structure in case of FeTe transforms into a prominent peak with a 
weak shoulder in case of FeSe. This could be 
associated with the changes in the tetragonal 
crystal structure of Fe(Se,Te). Substitution of Se for Te in FeTe leads to an 
increase in the Se/Te - Fe - Se/Te bond angle ($\alpha$ shown in fig 1 (a)). 
The angle $\alpha$ which is ~95$^{\circ}$ in case of FeTe, approaches the 
ideal tetrahedron value of 109.5$^{\circ}$ with Se doping\cite{yin}. 
The bond angle is determinant to 
the overlap between the iron and chalcogen 
orbitals, resulting in a stronger hybridization between the Fe 3d and the 
chalcogen p orbitals.

In Figure 1(e) we have plotted the valence band spectra from the three 
compositions taken at room temperature. As we see, with increase in the Se 
content from 0 to 1 the intensity of peak A increases. It can also be 
seen that, corresponding to this change in intensity some of the electronic 
states at the E$_F$ get depleted and the spectral weight at the E$_F$ shift 
to higher binding energy positions at A. Such a shift in spectral weight as 
a function of doping was observed earlier by{\it Yokoya et al}\cite{yokoya}. 
More importantly, the feature A shows an increase in its intensity as we go 
from 300 K to 77 K. This change is temperature induced. Although, in case of 
FeTe this increase is very small, substitution of half of Te with Se results 
in a marked change. It should be noted that with further increase in Se 
content this enhancement of intensity of peak A becomes weaker, though still 
distinct. Associated with this increase in intensity there is a depletion of 
states at the Fermi level. Further, it can be seen that the area by which the 
A peak has increased does not match with the number of states depleted from 
the near E$_F$ position. This indicates that electrons from other orbitals 
also shift resulting in the increase in its intensity. It should be noted 
that, depletion of these states from the near E$_F$ clearly indicates an 
opening up of a pseudogap as the temperature is lowered from 300 K to 77 K.

Figure 2(a) shows the calculated total density of states (DOS) over the valence 
band region. The calculated DOS matches with the earlier theoretical 
study\cite{miyake,subedi} and also with the observed experimental data. 
Features A and B exhibit predominant Fe 3d character, C represents hybridized 
Fe 3d and Se/Te p states while D corresponds to Se 4p/Te 5p states. Se 
incorporation shifts feature C and D towards higher binding energy owing to 
the greater electronegativity of Se (2.4) in comparison to Te (2.1). For 
FeTe, the features B and C merge while in case of FeSe, a clear gap is seen 
in the calculated DOS, which is in accordance with the experimental results.
The features marked A$^{\prime}$ (0.15 eV) and A$^{\prime\prime}$ (1.0 eV) 
(shown in Fig 2) corresponds to the experimentally obtained 
features at 0.1 eV and 0.5 eV, respectively. The discrepancy in the 
energy positions could be due to the self-energy correction which is neglected in the 
calculations\cite{yamasaki}.
The partial density of states (PDOS) for Fe 3d 
orbitals for an energy range 0 - 2.5 eV is represented in Fig. 2(b). 
 The PDOS (fig 2(b)) reveals that xz/yz and 
x$^{2}$-y$^{2}$ orbitals are the most populated ones at 0.15 eV (A$^{\prime}$) 
while the feature at 1 eV (A$^{\prime\prime}$) corresponds to states 
mostly populated with 3z$^{2}$-r$^{2}$. Although, FeSe, FeTe and 
FeSe$_{0.5}$Te$_{0.5}$ have tetragonal symmetry, different Fe-Se and Fe-Te 
bond lengths in case of FeSe$_{0.5}$Te$_{0.5}$ reduce the space group symmetry 
to 99 (P4mm)\cite{louca}. The lower symmetry in case of FeSe$_{0.5}$Te$_{0.5}$ results 
in the lifting of the degeneracy of xz/yz orbitals (as seen from fig 2(b)). 
The feature at $\sim$ 0.15 eV shifts back to higher binding energy and becomes 
less prominent with doping of Se in place of Te. On the contrary, the feature 
at $\sim$ 1.0 eV shifts towards E$_{F}$ and the plateau at $\sim$0.8 eV reduces 
with Se incorporation. The above effect results in an enhanced DOS for FeSe 
in comparison to FeTe at 0.5 eV (experimentally). The change in the orbital 
contribution is related to the chalcogen height from the Fe plane. Replacement 
of Te by Se leads to decrease in the chalcogen height, which affects the out 
of plane d orbitals, 3z$^{2}$-r$^{2}$ and xz/yz. The reduced chalcogen height 
leads to greater orbital overlap of 3z$^{2}$-r$^{2}$ and xz/yz orbitals with 
x$^{2}$-y$^{2}$ and xy orbitals, respectively, which in turn results in an 
orbital selective spectral weight transfer as seen from fig 3. The occupancy 
of xz/yz and x$^{2}$-y$^{2}$ orbitals reduces at E$_{F}$ with a simultaneous 
increase in occupancy of 3z$^{2}$-r$^{2}$ orbitals at higher binding energy. 
This change in the orbital occupancy is reflected as the spectral weight 
shifts which in turn results in the formation of a pseudogap with the doping 
of Se. It should be noted from fig 1(c) that the spectral weight transfer 
is weak in case of FeTe while FeSe$_{0.5}$Te$_{0.5}$ exhibits a 
significant shift.

A comparison of the figures 1(d) and 1(e) will reveal that the doping 
dependent spectral weight transfer is similar to the temperature dependent 
transfer in their energy positions. It has been shown earlier that lowering 
the temperature results in the reduction in the Fe-chalcogen height (Z shown 
in fig 1(a)) in case of FeTe and FeSe$_{0.5}$Te$_{0.5}$ while a slight 
increase in case of FeSe\cite{horigane}. The magnitude of this decrease is 
greater in case of FeSe$_{0.5}$Te$_{0.5}$ compared to FeTe. Thus, at 77 K, 
FeSe$_{0.5}$Te$_{0.5}$ has the shortest chalcogen height while FeTe the 
longest. The reduced chalcogen height at low temperature in case of 
FeSe$_{0.5}$Te$_{0.5}$ ensues the strongest hybridization between the orbitals.
This reflects in the maximum spectral weight transfer to higher binding 
energy in comparison to both the parent compounds. As mentioned before, in 
an analogy with the doping dependent case discussed earlier, the reduction 
in the Fe-Chalcogen height shifts the electron occupancy from the xz/yz and 
x$^{2}$-y$^{2}$ to the 3z$^{2}$-r$^{2}$ orbitals resulting in the temperature 
dependent spectral weight transfer and thereby the pseudogap. This is a 
temperature induced crossover from a metallic state in which all the t$_{2g}$ 
orbitals (xy, xz/yz) near the E$_F$ are occupied to a state in which 
occupancy of the xz/yz orbitals are depleted, signifying a Mott transition. 
This kind of spectral weight transfer which is a characteristic of 
Mottness\cite{phillips}, has earlier been identified as Orbital Selective 
Mott Transition (OSMT)\cite{medici}. Our study presents the first observation 
of a temperature induced crossover to a low temperature OSM phase in the 
family of Fe chalcogenides although such phenomena was earlier observed in 
A$_{x}$Fe$_{2-y}$Se$_{2}$ (A = K, Rb) where a high temperature OSM phase was 
identified\cite{yi}. Such spectral weight redistributions with lowering of 
temperature were earlier ascribed to the spin density wave transition (SDW) 
in FeTe by {\it Zhang et al.}\cite{zhang} who concluded that the suppression of 
the SDW is the cause of the reduction in the near E$_F$ spectral weight at 
low temperatures. But, our observation of a stronger spectral weight shift 
in case of FeSe$_{0.5}$Te$_{0.5}$ compared to that of FeTe shows that the 
spectral weight shift in this case is not related to the SDW transition. A 
recent ARPES study \cite{ieki} has reported the evolution of the spectral 
feature (a hump which corresponds to feature A in our data) as a function of 
x. The intensity of the hump was found to reduce with decreasing Se content. 
This study points at the role of electronic correlations driving the system 
close to the Mott metal insulator transition. Another ARPES study, temperature 
dependent, by {\it Liu et al.} \cite{liu} have reported a peak-dip-hump line 
shape in case of Fe$_{1.02}$Te across its antiferromagnetic transition at 70 
K. This study has shown the hump to become broader as the temperature is 
lowered below 50K and explained the results in terms of the strength of 
polarons. It should be noted that the spectral weight shifts observed by us 
are all above the Neel temperature. Nevertheless, these results 
together highlight the intricate electron correlation in the near E$_F$ 
states over a broad range of temperature. In Figure 4 we have compared the 
spectra collected using the HeI and HeII photons for FeTe, 
FeTe$_{0.5}$Se$_{0.5}$ and FeSe respectively. The black 
and red spectra correspond to data taken at 300 K and 77 K respectively. 
There is an enhancement of Fe 3d derived states in case of He II for all 
the compositions due to the higher cross section of Fe 3d at the He II energy.
The spectral weight shifts and thereby arising pseudogap described above 
follow the same trend in case of He II also but with a lower magnitude 
compared to He I.

In conclusion, we have studied the temperature and doping dependent spectral 
changes in the near E$_F$ valence states in FeSe$_{1-x}$Te$_{x}$. The strong 
orbital dependent spectral weight transfer at low temperature suggests that 
these compounds are in close proximity with Mottness. Using theoretical band 
structure calculations we have shown that the spectral weight transfer is due 
to the shifting of the electron occupancy from the xz/yz and x$^{2}$-y$^{2}$ 
orbitals to the 3z$^{2}$-r$^{2}$ indicating a temperature induced crossover 
from a metallic state to an Orbital Selective Mott (OSM) Phase.

\newpage
{\begin{figure}
\centering
\includegraphics[width=8cm]{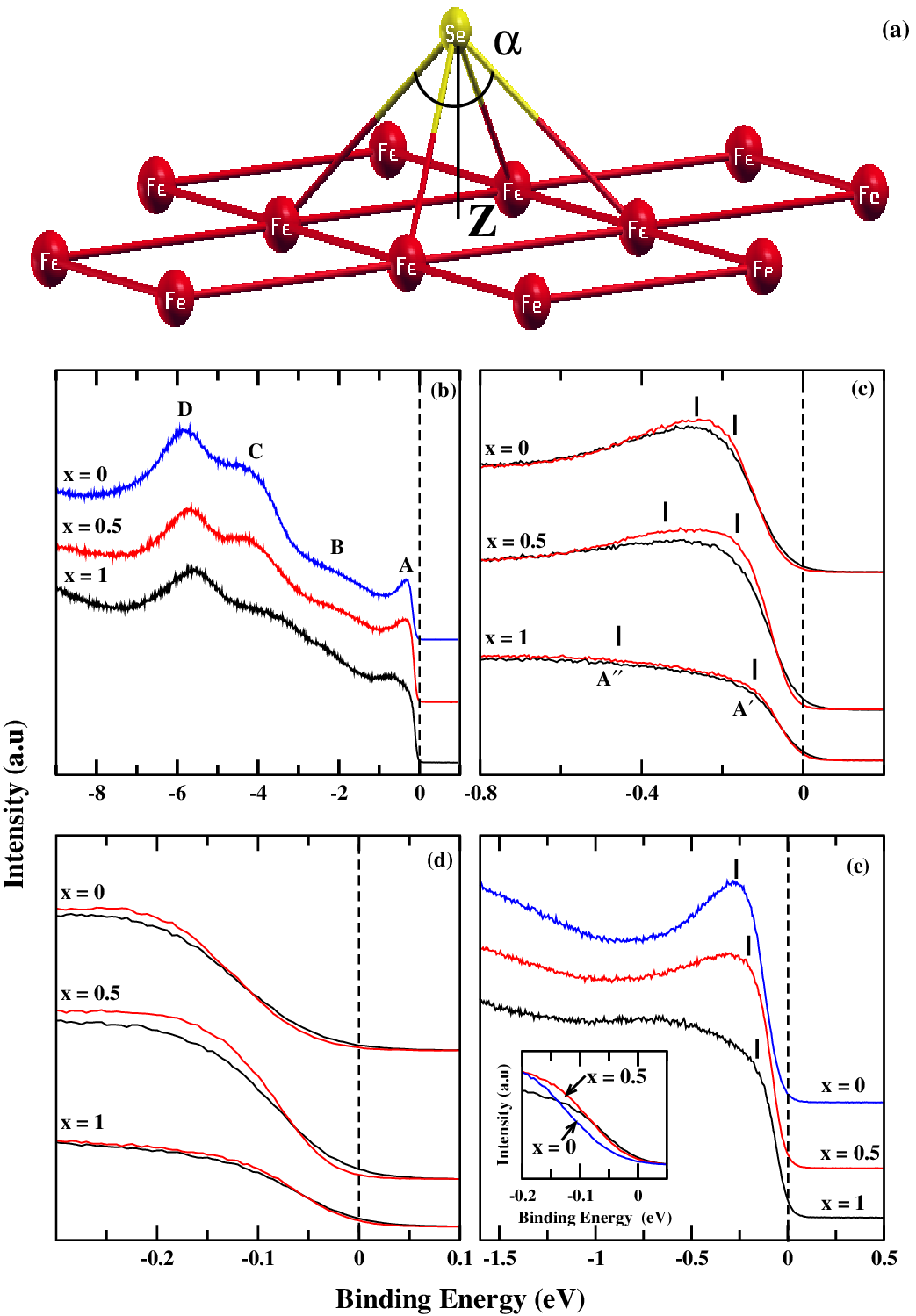}
\caption{(a) Structure of Fe(Se,Te)$_{4}$ tetrahedra showing the chalcogen 
height Z and angle $\alpha$. (b) He I Valence band spectra measured at 300 K for 
FeSe$_{1-x}$Te$_{x}$ (x = 1, 0.5, 0). The features are marked as A, B, C and D. (c) High resolution 
He I spectra measured at 300 K (black) and 77 K (red). Features A$^{\prime}$ and 
A$^{\prime\prime}$ are denoted by black bars. (d) Enlarged view 
of Near E$_{F}$ spectra showing the pseudogap. (e) High resolution He I 
spectra at 300 K. Inset shows the expanded view near the E$_{F}$.} 
\end{figure}}
{\begin{figure}
\centering
\includegraphics[width=10cm]{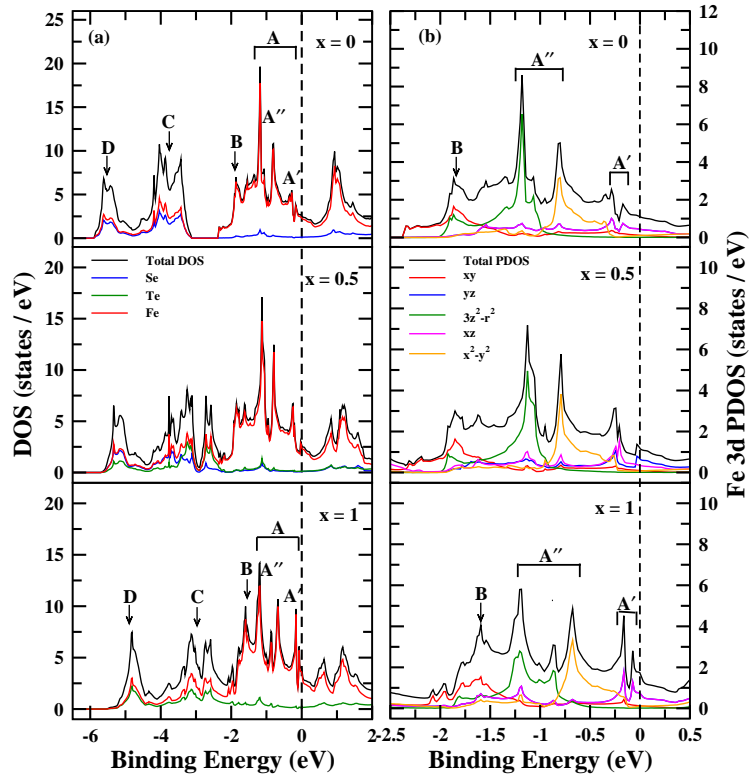}
\caption{Theoretically calculated (a) Total Density of States (DOS) 
(b) Fe 3d Partial DOS, using LDA+U, for FeSe$_{1-x}$Te$_{x}$ (x = 1, 0.5, 0). 
The theoretical features which matches with experiment has been marked as A, B, C and D.
A comprises of two features A$^{\prime}$ and A$^{\prime\prime}$ which are more 
clearly visible in panel (b). The PDOS contribution (panel b)
for xz (magenta) and yz (blue) orbitals are same for FeTe and FeSe. For 
FeSe$_{0.5}$Te$_{0.5}$, the xz and yz PDOS split owing to the different bond 
length of Fe-Te and Fe-Se.}
\end{figure}}
{\begin{figure}
\centering
\includegraphics[width=10cm]{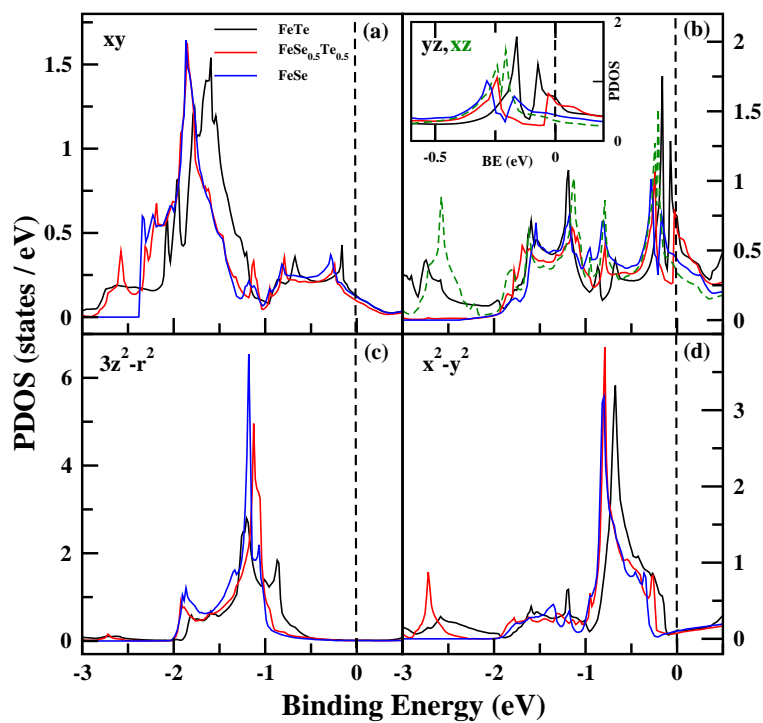}
\caption{The individual contributions of (a) xy, (b) yz/xz, (c) 3z$^{2}$-r$^{2}$
and (d) x$^{2}$-y$^{2}$ orbitals to Fe 3d partial DOS for FeTe, FeSe$_{0.5}$Te$_{0.5}$
and FeSe. Inset of (b) represents the expanded PDOS near E$_{F}$ for yz (solid lines) 
and xz (dotted green line, for FeSe$_{0.5}$Te$_{0.5}$) orbitals. The reduced 
symmetry in case of FeSe$_{0.5}$Te$_{0.5}$  lifts the degeneracy of xz and yz orbitals, 
thus the PDOS contribution splits which is otherwise same in case of FeTe and FeSe.}
\end{figure}}
{\begin{figure}
\centering
\includegraphics[width=8cm]{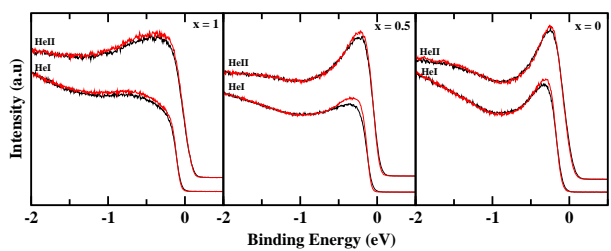}
\caption{Comparasion of HeI and HeII spectra for FeSe$_{1-x}$Te$_{x}$ (x = 1, 0.5, 0)
measured at 300 K (black) and 77 K (red).}
\end{figure}}

\end{document}